\documentclass[conference]{IEEEtran}
\IEEEoverridecommandlockouts

\def\BibTeX{{\rm B\kern-.05em{\sc i\kern-.025em b}\kern-.08em
    T\kern-.1667em\lower.7ex\hbox{E}\kern-.125emX}}
\usepackage{cite}

\ifCLASSINFOpdf
   \usepackage[pdftex]{graphicx}
  \graphicspath{{./Figs/}}
   \DeclareGraphicsExtensions{.pdf,.jpg,.JPG,.png,.eps}
\else
   \usepackage[dvips]{graphicx}
   \graphicspath{{./Figs/}}
   \DeclareGraphicsExtensions{.eps}
\fi
\usepackage{epstopdf}

\usepackage{amsmath}
\usepackage{algorithm,algorithmic}

\usepackage{multirow}
\usepackage{pbox}
\usepackage{array}

\ifCLASSOPTIONcompsoc
  \usepackage[caption=false,font=normalsize,labelfont=sf,textfont=sf]{subfig}
\else
 \usepackage[caption=false,font=footnotesize]{subfig}
\fi
\usepackage{dblfloatfix}

\usepackage[hyphens]{url}
\usepackage[hidelinks,bookmarks=false]{hyperref}
\hypersetup{breaklinks=true}
\usepackage{wasysym}
\usepackage{bm}
\usepackage{amssymb}
\usepackage{gensymb}
\usepackage{flushend}
\usepackage{makecell}
\usepackage{siunitx}
\usepackage{fixltx2e}
\usepackage{xspace}
\usepackage{datetime}
\usepackage[dvipsnames]{xcolor}

\newcommand{\RAunits}{\si[inter-unit-product = \ensuremath{\cdot}]{\ohm\um\squared}\xspace}
\usepackage{doi}

\begin{document}
	
\bstctlcite{IEEEexample:BSTcontrol}
%

\title{Impact of  Magnetic Coupling and  Density\protect\\ on STT-MRAM Performance \vspace{-5mm}}
 \author{
 	\IEEEauthorblockN{\,Lizhou Wu\IEEEauthorrefmark{1} {}   Siddharth Rao\IEEEauthorrefmark{3} {}   Mottaqiallah Taouil\IEEEauthorrefmark{1}\IEEEauthorrefmark{2} {}     Erik Jan Marinissen\IEEEauthorrefmark{3} {}    Gouri Sankar Kar\IEEEauthorrefmark{3}  {}   Said Hamdioui\IEEEauthorrefmark{1}\IEEEauthorrefmark{2}}
 	\IEEEauthorblockA{ \IEEEauthorrefmark{1}TUDelft, Delft, The Netherlands {} {}{} {}{} {} {}  {}{} {} {}   \IEEEauthorrefmark{2} CognitiveIC, Delft, The Netherlands {} {} {}{} {}{} {}  {} {}{} {} \IEEEauthorrefmark{3}IMEC, Leuven, Belgium 
 		\\ \{Lizhou.Wu, M.Taouil, S.Hamdioui\}@tudelft.nl    {} {}  {} {} {} {} {} {}    {} {} {} {} {} {} {} {} {} {} \{Siddharth.Rao, Erik.Jan.Marinissen, Gouri.Kar\}@imec.be}
     \vspace{-10mm}
 }

\maketitle
\begin{abstract}

As  a unique mechanism  for MRAMs, magnetic coupling needs to be accounted for when designing memory arrays.
This paper models both intra- and inter-cell magnetic coupling analytically for STT-MRAMs and investigates their impact on the write performance and retention of  MTJ devices, which are the data-storing elements of STT-MRAMs.
We present magnetic measurement data of MTJ devices with diameters ranging from 35\,nm to 175\,nm, which we use to calibrate our intra-cell magnetic coupling model.  Subsequently, we extrapolate this model to study inter-cell magnetic coupling in memory arrays. We propose  the inter-cell magnetic coupling factor $\bm{\Psi}$ to indicate  coupling strength. Our simulation results show that $\bm{\Psi}$$\bm{\approx}$$\bm{2\%}$ maximizes the array density under the constraint that the magnetic coupling has negligible impact on the device's  performance. Higher array densities show significant variations in average switching time, especially at low switching voltages, caused by inter-cell magnetic coupling, and dependent on the data pattern in the cell's  neighborhood. We also observe a marginal degradation of the data retention time under the influence of inter-cell magnetic coupling.
\end{abstract}


%
%

%
\IEEEpeerreviewmaketitle

\vspace*{-6pt}
\section{Introduction}
\label{sec:introduction}
Spin-transfer torque magnetic random access memory (STT-MRAM) is considered as one of the most promising  non-volatile memory technologies, since it features high density, nearly unlimited endurance, and negligible leakage power \cite{Fong2016}.
Thus,  many foundries worldwide have been  investing heavily in its commercialization.
For example, SK hynix demonstrated in 2016 \cite{Chung2016} a 4\,Gb STT-MRAM prototype  targeting the replacement of DRAM and flash memories.
Samsung and Intel also  presented their  STT-MRAM solutions in 2018 \cite{SongYJ2018} and 2019 \cite{Wei2019a}, respectively. 
In STT-MRAMs, data is stored in magnetic tunnel junction (MTJ) devices which contain multiple ferromagnetic layers. Each of them generates a stray field, which has a significant impact on the device's performance.  It has been shown that the stray field increases  as the MTJ dimension shrinks \cite{Han2015, Wu2020survey}, which makes \textit{intra-cell magnetic coupling} a critical constraint for STT-MRAM designs at advanced technology nodes. Furthermore, to compete with DRAM and flash memories, high-density STT-MRAM arrays are  required. It was reported that the STT-MRAM  array pitch can be made as small as 1.5$\times$ the MTJ diameter at sub-\SI{20}{\nano\meter} nodes, using advanced nano-patterning techniques  \cite{Nguyen2018}. As the pitch decreases, MTJ devices are pushed closer to each other. This makes \textit{inter-cell magnetic coupling} between neighboring cells  become increasingly evident, which may lead to write errors \cite{chappert2010emergence}. Therefore, it is crucial to analyze magnetic coupling and evaluate its impacts quantitatively when it comes to high-density STT-MRAMs at advanced technology  nodes.

There is limited work published on  magnetic coupling considering both intra- and inter-cell effects in dense STT-MRAM arrays  based on  designs that were  manufactured. 
In \cite{Jiancheng2015}, Huang et al.   observed experimentally that the stray field results in a distinct difference in the thermal stability factor $\Delta$ of the two binary states of MTJ devices. 
Wang et al. \cite{Wang2012} reported that the stray field has a non-uniform distribution  over the cross-section of the MTJ device, and it results in a significant variation in the switching time based on micromagnetic simulations. In \cite{Golonzka2018}, Golonzka et al. even observed that some of  Intel's devices were  locked to one certain state due to  a strong stray field. However, these papers did not take into consideration  the stray fields from neighboring cells. Augustine et al. \cite{Augustine2010} analyzed the stray field from four adjacent cells for in-plane MTJ devices and concluded that inter-cell  stray fields can cause up to 80\% increase in  switching time. Yoon et al. \cite{Yoon2018} explored the effect of inter-cell magnetic coupling on the MTJ's  properties in a compact memory array. Overall, the related prior work has  the following shortcomings: 1) not based on real-world MTJ stack designs with  perpendicular magnetic anisotropy, dual MgO, and synthetic anti-ferromagnetic (SAF) pinned layer;  2)  intra- and inter-cell magnetic coupling are not considered simultaneously; 3) a lack of magnetic characterization data.

In this paper, we address these issues by characterizing and subsequently modeling intra-cell magnetic coupling for isolated MTJ devices fabricated at IMEC.  This model is then used to calculate stray fields at the victim cell located in the center of a  representative 3$\times$3 memory array. The  contributions of this paper are as follows.
\vspace*{-2pt}
\begin{itemize}
   \item Magnetic characterization results of MTJ devices with various sizes ranging from \SI{35}{\nano\meter} to \SI{175}{\nano\meter}.
   \item An analytical intra-cell magnetic coupling model, which is calibrated and validated by silicon data.
   \item Extrapolating the above model to study inter-cell magnetic coupling on a  cell array with varying pitches.
    \item Introduction of the inter-cell magnetic coupling factor $\Psi$ to indicate the coupling strength.
   
   \item Evaluation of the impact of magnetic coupling on the  MTJ's  write characteristics and  retention time.
   
\end{itemize}

The rest of this paper is organized as follows.
Section~\ref{sec:background} provides a background on the MTJ device technology and the  magnetic coupling mechanism.
Section~\ref{sec:intracell_mc_characterization} presents  magnetic characterization data.
Section~\ref{sec: MC_modeling} elaborates the modeling methodology of intra-cell and inter-cell magnetic coupling.
Section~\ref{sec:MC_impact} evaluates the impact of magnetic coupling on the MTJ's performance. 
 Section~\ref{sec:conclusion}   concludes this paper.

\section{Background}
\label{sec:background}


\subsection{MTJ Device Technology}
\begin{figure}[!t]
	\centering
	\includegraphics[width=0.5 \textwidth ]{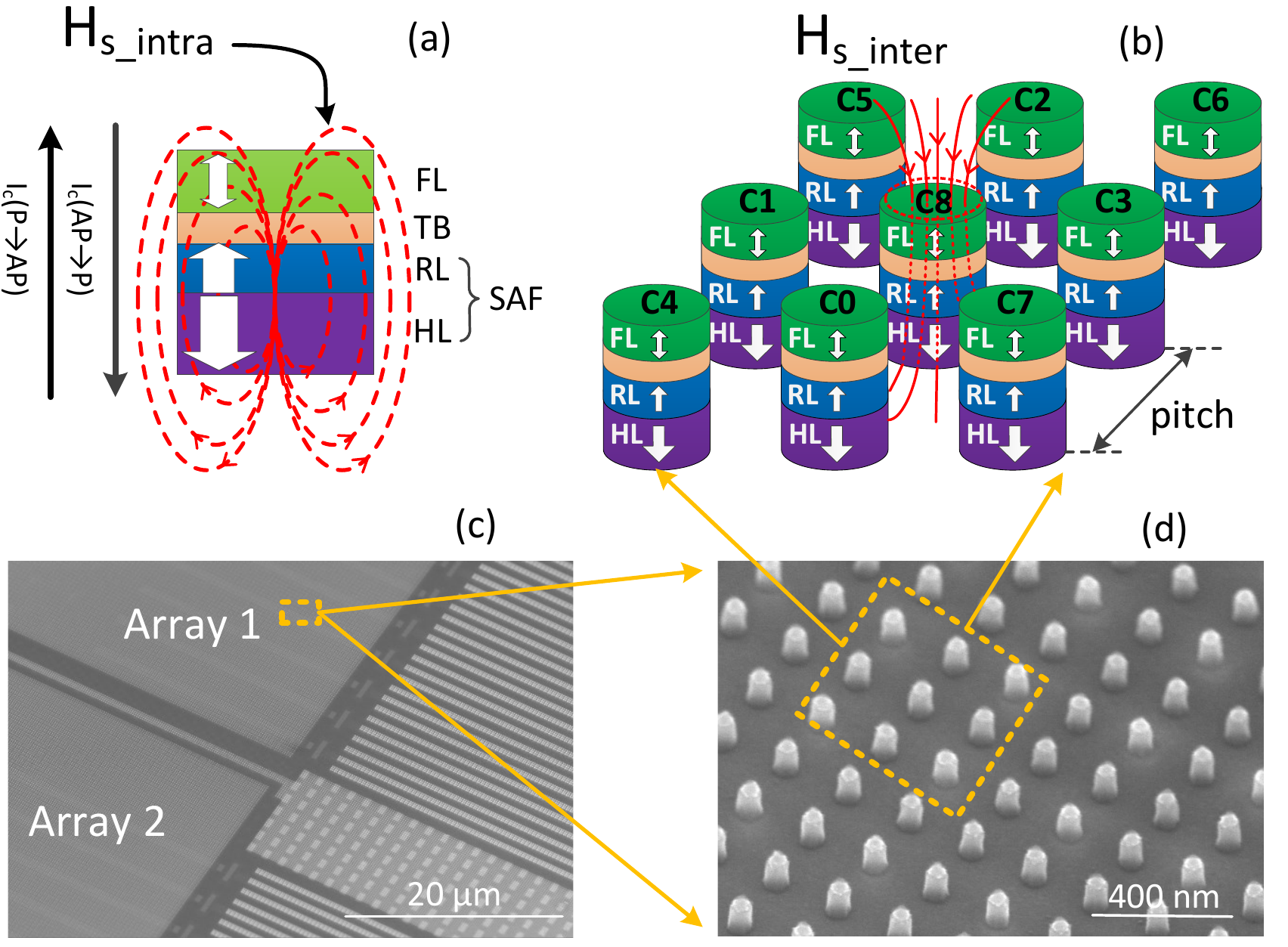}
	\caption{ (a)  MTJ stack and the intra-cell stray fields from the RL and HL, (b) 3$\times$3 MTJ array and the inter-cell stray fields from neighboring cells, (c)  SEM image of the 0T0R wafer floor plan, and (d) SEM image of MTJ array.}
	\label{fig:STTMRAM_basics}
\end{figure}

{\em Magnetic tunnel junction} (MTJ) devices are the data-storing elements in STT-MRAMs. 
Each MTJ device stores one-bit data in the form of binary magnetic configurations \cite{Wu2019TETC}. Fig.~\ref{fig:STTMRAM_basics}a shows the MTJ stack which essentially consists of four layers: FL/TB/RL/HL. 
The {\em hard layer} (HL) is composed of [Co/Pt]\textsubscript{{x}}, which is used to pin the magnetization in the upper {\em reference layer} (RL).  The RL is generally built up with a Co/spacer/CoFeB multilayer, which is anti-ferromagnetically coupled to the HL. These two layers form a {\em synthetic anti-ferromagnetic} (SAF) structure, providing a strong fixed reference magnetization in the RL. The {\em tunnel barrier} (TB) layer is made of dielectric MgO, typically $\sim$\SI{1}{\nano\meter}. The {\em resistance-area} (RA) product is commonly used to evaluate the TB resistivity, as it depends on the TB thickness but not the device size.
The CoFeB-based {\em free layer} (FL) is the  data-storing layer where the magnetization can be switched by a spin-polarized current. 

To work properly as  memory devices, MTJs need to provide read and write mechanisms, which are realized by the {\em tunneling magneto-resistance} (TMR) effect  and the {\em spin-transfer-torque} (STT) effect \cite{ Khvalkovskiy2013d}. Due to the TMR effect, the MTJ's resistance is low ($R_{\mathrm{P}}$) when the magnetization in the FL is parallel  to that in the RL; the resistance is high ($R_{\mathrm{AP}}$) when in anti-parallel  state. 
If the write current magnitude (with sufficiently long pulse width) is larger than the {\em critical switching current} ($I_{\mathrm{c}}$), the magnetization in the FL can switch to the opposite direction.  It is a fundamental parameter to characterize the switching capability by current. The STT-induced switching behavior also depends on the current direction, as shown in Fig.~\ref{fig:STTMRAM_basics}a. $I_{\mathrm{c}}$(AP$\rightarrow$P) can be significantly different from  $I_{\mathrm{c}}$(P$\rightarrow$AP) due to the bias dependence of STT efficiency and external field disturbance \cite{Khvalkovskiy2013d}. In addition, the {\em average switching time} ($t_{\mathrm{w}}$) \cite{Wu2018} is another critical parameter, which is inversely correlated with the  write current. In other words, the higher the write current over $I_{\mathrm{c}}$, the less the time required for the magnetization in FL to flip. In practice, $t_{\mathrm{w}}$(AP$\rightarrow$P) can also differ from $t_{\mathrm{w}}$(P$\rightarrow$AP) depending on the  write current magnitude and duration.

In addition, enough retention time is required for STT-MRAMs depending on the target application.  Storage applications require  $>$10\,years typically, while cache applications only necessitate ms-scale retention time \cite{Jog2012a}.   an STT-MRAM retention fault occurs when the magnetization in the FL of the MTJ flips spontaneously due to thermal fluctuation.  Thus, the STT-MRAM retention time is generally characterized by the {\em thermal stability factor} ($\Delta$) \cite{Khvalkovskiy2013d}. The higher the $\Delta$, the longer the retention time. 

\subsection{Magnetic Coupling Mechanism }
To obtain high TMR and strong interfacial perpendicular magnetic anisotropy (iPMA),  our MTJ devices were annealed at \SI{375}{\celsius} for \SI{30}{\minute} in a vacuum chamber under the perpendicular (out-of-plane) magnetic field of \SI{20}{\kilo Oe}. Once the ferromagnetic layers (i.e., FL, RL, and HL) in the MTJ stack are magnetized, each of them inevitably generates a  stray field in the space. Fig.~\ref{fig:STTMRAM_basics}a illustrates the intra-cell stray field ($\bm{H_{\mathrm{s\_intra}}}$) perceived at the FL, generated by the RL and HL together; its in-plane component ($H_{\mathrm{s\_intra}}^{x-y}$) is marginal \cite{Wang2012}, while its  out-of-plane component ($H_{\mathrm{s\_intra}}^{z}$)  at the FL has a significant influence on  the energy barrier ($E_{\mathrm{b}}$) between the P and AP states \cite{Fong2016}. For example, if  $H_{\mathrm{s\_intra}}^{z}$  has the same direction as the magnetization in the FL in AP state, it leads to an increase in $E_{\mathrm{b}}$(AP$\rightarrow$P) and a decrease in $E_{\mathrm{b}}$(P$\rightarrow$AP). The deviation of energy barriers along the two switching directions has a significant impact on the retention  and the STT-switching characteristics  of MTJ devices, as reported in \cite{Wang2012,Jiancheng2015,Golonzka2018}.
In the extreme case where $H_{\mathrm{s\_intra}}^{z}$ exceeds the FL {\em coercivity} ($H_{\mathrm{c}}$), defined as the reverse field needed to drive the magnetization of a ferromagnet to zero, the bistable states will disappear \cite{Han2015}.

Furthermore,  as the density of STT-MRAMs increases, the spacing between neighboring MTJ devices becomes narrower (i.e., smaller pitch). This makes stray fields from neighboring cells not negligible any more \cite{Augustine2010,Fong2016}. 
Fig.~\ref{fig:STTMRAM_basics}b shows a 3$\times$3 MTJ array, where the eight cells C0-C7 (aggressors) surrounding cell C8 (victim) in the center inevitably generate an inter-cell stray field ($\bm{H_{\mathrm{s\_inter}}}$) acting on the victim cell. Fig.~\ref{fig:STTMRAM_basics}c and Fig.~\ref{fig:STTMRAM_basics}d show the  scanning electron microscope (SEM)  images of our 0T1R wafer floor plan and MTJ array, respectively. 

\section{ Intra-Cell Magnetic Coupling Characterization}
\label{sec:intracell_mc_characterization}

\begin{figure}[!t]
	\centering
	\includegraphics[width=0.5 \textwidth ]{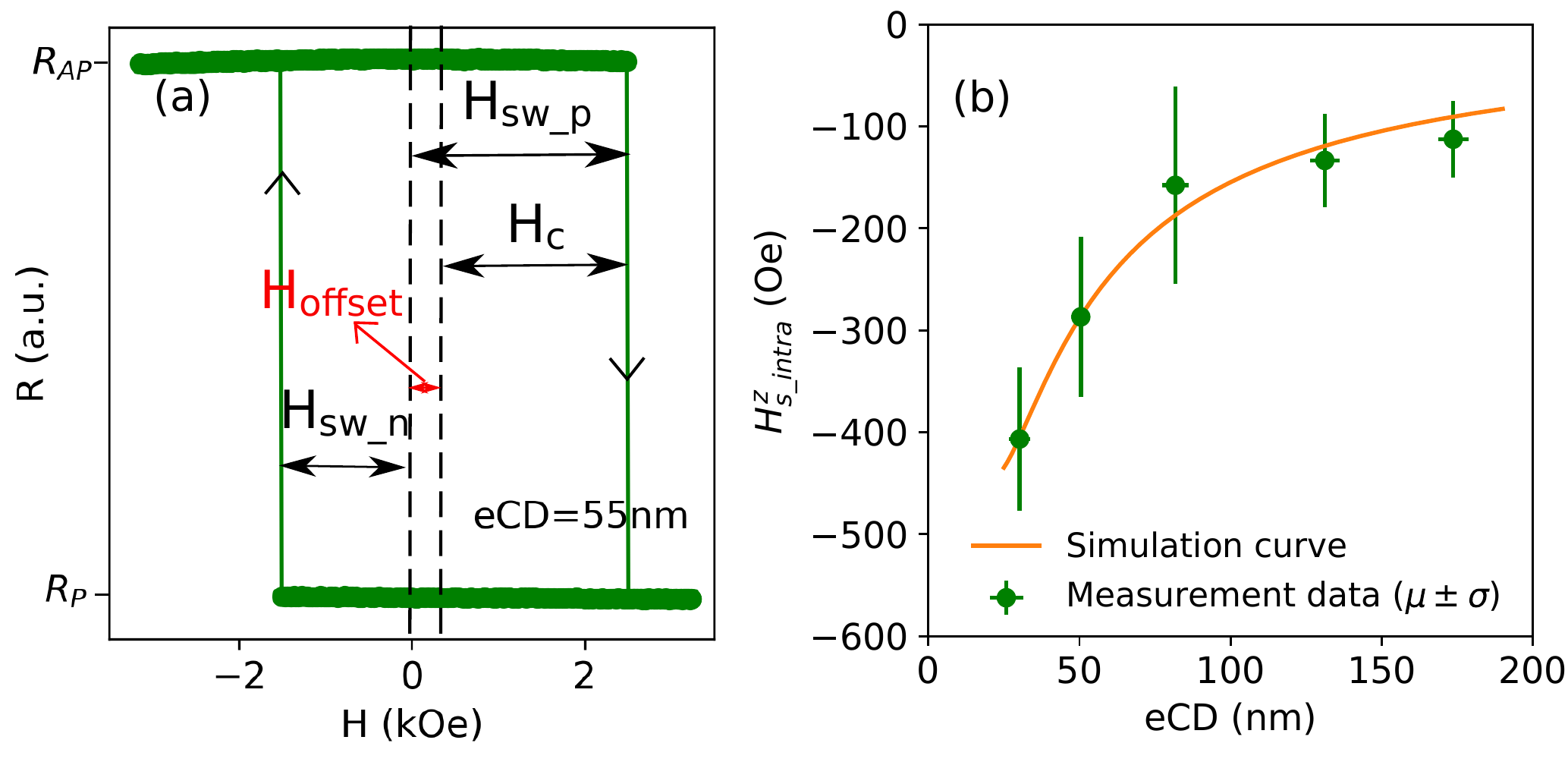}
	\caption{(a) Measured R-H hysteresis loop, (b) device size dependence of $H_{\mathrm{s\_intra}}^z$: measured vs. simulated.}
	\label{fig:Hs_characterization}
\end{figure}

 $H_{\mathrm{s\_intra}}^{z}$  can be extracted from R-H hysteresis loops. Fig.~\ref{fig:Hs_characterization}a shows a  measured  R-H hysteresis loop for a representative MTJ device with the HL/RL configuration shown in Fig.~\ref{fig:STTMRAM_basics}a.
During the measurement, an external field was applied perpendicularly  to the device under test. It was ramped up from \SI{0}{Oe} to \SI{3}{kOe}, then it went backwards to \SI{-3}{kOe} and finished at \SI{0}{Oe}. In total, we measured 1000 field points, each of which was followed by a read  operation to read out the device resistance with a voltage of \SI{20}{\milli\volt}.  It can be seen that  the MTJ device switches from AP state (high resistance) to P state (low resistance) when the field reaches at $H_{\mathrm{sw\_p}}$, and it switches back to AP state at a negative field $H_{\mathrm{sw\_n}}$. The device coercivity can be obtained by $H_{\mathrm{c}}=(H_{\mathrm{sw\_p}}-H_{\mathrm{sw\_n}})/2$.  Due to the existence of  stray fields at the FL, the loop is always offset to the positive side for the device configuration  in Fig.~\ref{fig:STTMRAM_basics}a. The offset field $H_{\mathrm{offset}}$ is equal to $(H_{\mathrm{sw\_p}}+H_{\mathrm{sw\_n}})/2$, as shown in the figure. Since $H_{\mathrm{offset}}$ is essentially equivalent to the extra external field applied to cancel out $H_{\mathrm{s\_intra}}$, the relation of these two parameters is $H_{\mathrm{s\_intra}}=-H_{\mathrm{offset}}$. Given the fact that the {\em resistance-area product} ($\mathit{RA}$) does not change with the device size, the {\em electrical Critical Diameter} (eCD) of each device can be derived by \cite{JWu2018}:
$eCD=\sqrt{\frac{4}{\pi}\cdot \frac{RA}{R_{\mathrm{P}}}},$
where $\mathit{RA}$$=$$4.5$\,\RAunits (measured at blanket stage) for this wafer, and $R_{\mathrm{P}}$ can be extracted from the R-H loop (i.e., the lower horizontal line in Fig.\ref{fig:Hs_characterization}a). The calculated eCD=\SI{55}{\nano\meter} for the device shown in Fig.\ref{fig:Hs_characterization}a.

In this way, we can obtain $H_{\mathrm{s\_intra}}^{z}$ and eCD for MTJ devices with different sizes on the same wafer.  The measurement results are shown in Fig. \ref{fig:Hs_characterization}b. The  error bars indicate the device-to-device variation in the the measured values due to process variations and the intrinsic switching stochasticity.    
It can be seen that the smaller the device size (i.e., smaller eCD), the higher $H_{\mathrm{s\_intra}}^{z}$; the trend even tends  to grow exponentially for eCD$<$\SI{100}{\nano\meter}. The solid curve in the figure represents simulation results which will be explained in the next section.
\section{Modeling of Magnetic Coupling}
\label{sec: MC_modeling}
To analyze the impact of magnetic coupling on the MTJ's performance, we need to develop an analytical model for both inter-cell and intra-cell magnetic coupling. In this section, we first model and calibrate $\bm{H_{\mathrm{s\_intra}}}$ for isolated MTJ devices, based on  the measurement data as presented in the previous section. Thereafter, we extrapolate this model to derive  $\bm{H_{\mathrm{s\_inter}}}$ for an  memory array with various pitches.

\subsection{Intra-Cell Magnetic Coupling }
\label{subsec:Intra_MC_modeling}

\begin{figure*}[!h]
	\centering
	\includegraphics[width=1 \textwidth ]{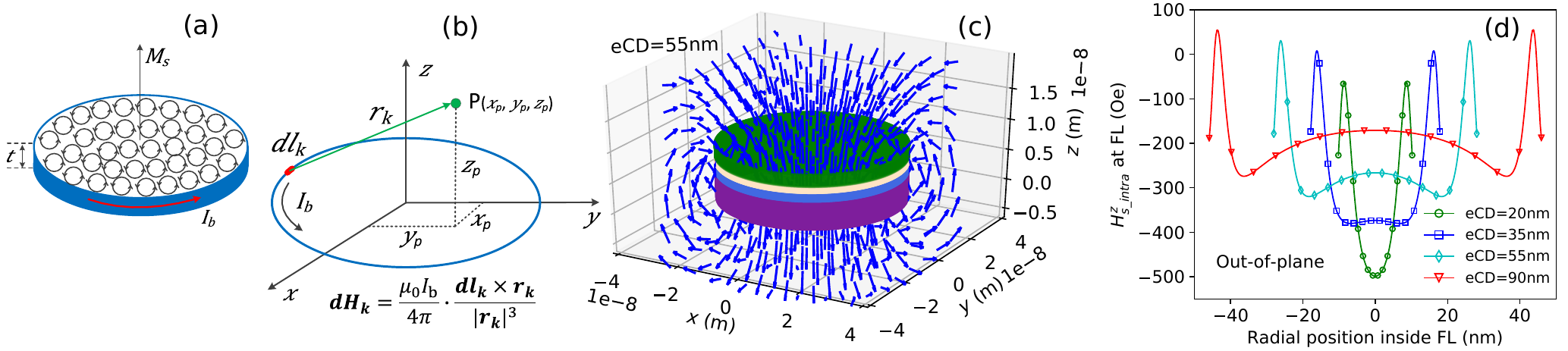}
	\caption{(a) Bound current, (b) Biot-Savart law, (c) intra-cell stray field $\bm{H_{\mathrm{s\_intra}}}$ from the HL and RL for an MTJ  with eCD=\SI{55}{\nano\meter}, and (d) the out-of-plane component $H_{\mathrm{s\_intra}}^z$ distribution over the cross-section of the FL, with respect to various eCDs.}
	\label{MC_modeling_results1}
\end{figure*}

Under the assumption that each ferromagnetic layer (i.e., FL, RL, and HL) in MTJ devices is uniformly magnetized, the produced field is identical to the field that would be produced by the bound current \cite{griffiths2013}.  Fig. \ref{MC_modeling_results1}a depicts a thin ferromagnet with tiny current loops representing dipoles. All internal currents cancel  each other while there is no adjacent loop at the edge to do the canceling. As a result, the net effect is a macroscopic current $I_{\mathrm{b}}$ (referred to as bound current) flowing around the boundary. The magnetic moment of this ferromagnet can be expressed as $\bm{m}  =\bm{M_s}\cdot A \cdot t $ \cite{griffiths2013},
where $\bm{M_s}$ is the saturation magnetization, $A$  is the cross-sectional area, and $t$ is the thickness of this ferromagnet.  Considering the bound current $I_{\mathrm{b}}$, $\bm{m}$ can also be written as $I_{\mathrm{b}}\cdot A \cdot \bm{\hat{n}}$
where $\bm{\hat{n}}$ is the unit vector along the direction of $\bm{M_s}$ \cite{griffiths2013}. Therefore, one can easily derive $I_{\mathrm{b}}=M_s t $. For each ferromagnet in the MTJ stack, the $M_st $ product is measured at blanket film level by vibrating sample magnetometry (VSM) measurements. 

With the derived bound current $I_{\mathrm{b}}$ for each ferromagnet in the MTJ stack, the generated stray field in the space can be modeled as the field of a current loop with current $I_{\mathrm{b}}$, as shown in Fig.~\ref{MC_modeling_results1}b. In this way, the stray field  at any point P($x_{\mathrm{p}}$,$y_{\mathrm{p}}$,$z_{\mathrm{p}}$) in the space can be calculated by the Biot-Savart law \cite{griffiths2013}:
\begin{equation}
	\bm{H}(\bm{r})=\frac{\mu_{0}}{4\pi}\oint \frac{I_{\mathrm{b}}\bm{dl}\times \bm{r}}{|\bm{r}|^3}, \label{BS_law}
\end{equation}
where $\bm{dl}$ is an infinitesimal length of the current loop,    $\bm{r}$  the vector distance from $\bm{dl}$ to the point P,  and $\mu_{0}$  the vacuum permeability.  To calculate the above integral in a discrete form, we can divide the current loop into a large number of small segments, thereafter sum up the fields of all segments at point P as an approximation of $\bm{H}(\bm{r})$.

Assume the current loop is cut into $N$ segments. For the $k^{th}$ segment  $\bm{dl_k}$ ($k\in[0,N-1]$),  we derive:
\begin{align*}
\bm{dl_k}&=(x_{k+1}-x_{k}, y_{k+1}-y_{k}, z_{k+1}-z_{k}), \\
\bm{r_k}&=(x_{p}-x_{k}, y_{p}-y_{k}, z_{p}-z_{k}).
\end{align*}
Therefore, $\bm{dl_k}\times \bm{r_k}=(S_k^x,S_k^y,S_k^z)$, where
\begin{align*}
	S_k^x&=(y_{k+1}-y_{k})\cdot(z_{p}-z_{k})-(z_{k+1}-z_{k})\cdot(y_{p}-y_{k}), \\
	S_k^y&=(z_{k+1}-z_{k})\cdot(x_p-x_k)+(x_{k+1}-x_k)\cdot(z_p-z_k), \\
	S_k^z&=(x_{k+1}-x_k)\cdot(y_p-y_k)-(y_{k+1}-y_k)\cdot(x_p-x_k).
\end{align*}
The field generated by the tiny segment $\bm{dl_k}$ is 
$$ \bm{dH_k}=(dH_k^x,dH_k^y,dH_k^z)=\frac{\mu_{0}}{4\pi} \cdot \frac{I_{\mathrm{b}}}{|\bm{r_k}|^3}\cdot(S_k^x,S_k^y,S_k^z).$$
By summing up the fields of all $N$ segments, we derive the overall  field at the spot P generated by the entire current loop: 
$$\bm{H}=\sum\limits_{k=0}^{N-1}\bm{dH_k} =(\sum\limits_{k=0}^{N-1}dH_k^x, \sum\limits_{k=0}^{N-1}dH_k^y, \sum\limits_{k=0}^{N-1}dH_k^z) $$

In this way, we can calculate the intra-cell stray field from the HL ($\bm{H_{\mathrm{s\_HL}}}$) and intra-cell stray field from the RL  ($\bm{H_{\mathrm{s\_RL}}}$), respectively. The overall intra-cell stray field is the vector sum of these two fields (i.e., $\bm{H_{\mathrm{s\_intra}}}=\bm{H_{\mathrm{s\_HL}}}+\bm{H_{\mathrm{s\_RL}}}$), which is visualized in Fig.~\ref{MC_modeling_results1}c for an MTJ device with eCD=\SI{55}{\nano\meter}. Fig. \ref{MC_modeling_results1}d shows the distribution of the z-component $H_{\mathrm{s\_intra}}^{z}$ (i.e., the out-of-plane component) over the horizontal cross-section of the FL. It can be seen that $H_{\mathrm{s\_intra}}^{z}$ is not uniformly distributed at the FL; its magnitude is smaller at the edge than at the center. We took the values at the center (i.e., at radial position=\SI{0}{\nano\meter}) and calibrated them with the measured data. Fig.~\ref{fig:Hs_characterization}b presents the simulation results of  $H_{\mathrm{s\_intra}}^{z}$ vs. eCD, which match the silicon data.


\subsection{Inter-Cell Magnetic Coupling  }
 \label{subsec:Inter_MC_modeling}

To study the inter-cell magnetic coupling effect, we extrapolate the  intra-cell magnetic coupling model from a single MTJ device to a  3$\times$3 MTJ array in Cartesian Coordinates. The nine devices are named  C0 to C8, as illustrated in Fig.~\ref{fig:STTMRAM_basics}b.  Cell C8 in the center is considered as  the victim  whereas the four direct neighbors (C0-C3) and four diagonal neighbors (C4-C7) are aggressor cells.  In this way, the inter-cell magnetic coupling effect is translated to the impact of net stray field from the eight neighboring cells (denoted as  $\bm{H_{\mathrm{s\_inter}}}$) on the FL of the victim C8.  $\bm{H_{\mathrm{s\_inter}}}$ can be calculated by:
$$ 
\bm{H_{\mathrm{s\_inter}}}=\sum_{i=0}^{7} (\bm{H_{\mathrm{s\_HL}}}(\mathrm{C}i)+\bm{H_{\mathrm{s\_RL}}}(\mathrm{C}i)+\bm{H_{\mathrm{s\_FL}}}(\mathrm{C}i)).$$
Since the HL and RL are both fixed layers after the fabrication of MTJ devices, $\bm{H_{\mathrm{s\_HL}}}$ and $\bm{H_{\mathrm{s\_RL}}}$ are fixed, given an eCD and a pitch node. However, the direction of  $\bm{H_{\mathrm{s\_FL}}}$ changes dynamically depending on the data stored in the MTJ device though its magnitude remains the same. As a result, $\bm{H_{\mathrm{s\_inter}}}$ depends on the {\em neighborhood pattern}  in the eight neighboring cells (i.e., C0-C7), which we denote as NP\textsubscript{8}. In the binary form, NP\textsubscript{8} can be expressed as:  
$[d_0,...,d_7]_2$,
where $d_i\in\{0,1\}$ represents the data stored in C$i$. In addition, NP\textsubscript{8} can also be transformed to the decimal form: $[n]_{10}$, where $n\in[0,255]$.

\begin{figure*}[!h]
	\centering
	\includegraphics[width=1 \textwidth ]{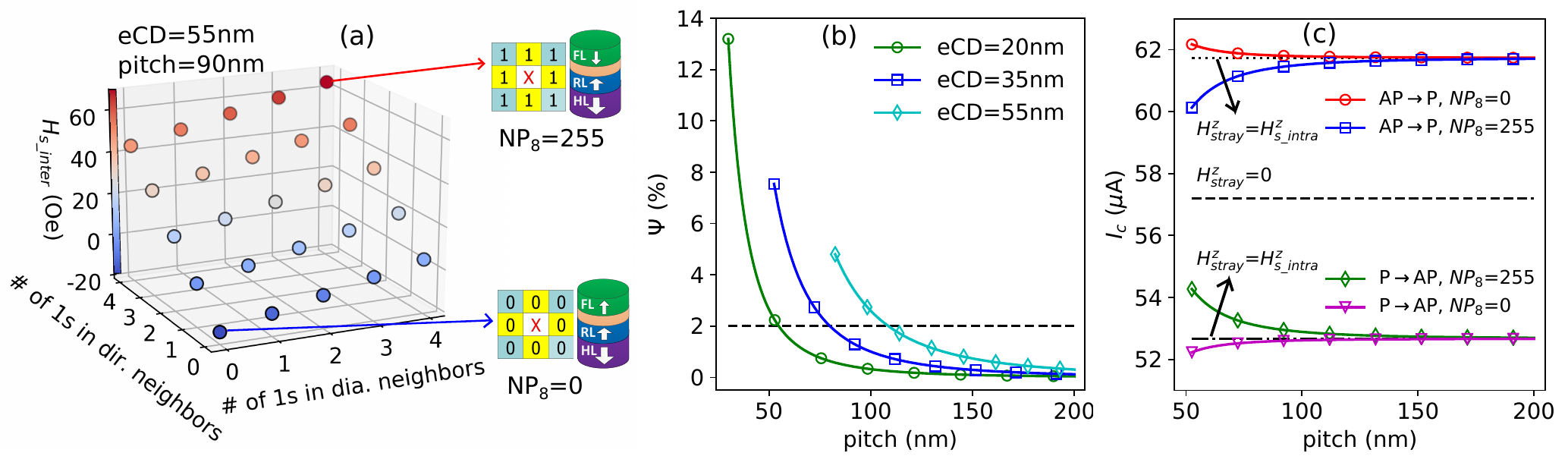}
	\caption{ (a) $H_{\mathrm{s\_inter}}^{z}$ at the FL of victim C8 under various combinations of the number of 1s in  direct neighbors and diagonal neighbors, (b) $\Psi$ vs. pitch with respect to three MTJ sizes, and (c) $I_{\mathrm{c}}$ vs. pitch under the circumstance of different stray fields. }
	\label{MC_modeling_results2}
\end{figure*}

Fig. \ref{MC_modeling_results2}a shows the resultant   $H_{\mathrm{s\_inter}}^z$ values at the FL of victim C8 as a function of the number of 1s in  direct neighbors C0-C3 (marked in yellow) and the number of 1s in diagonal neighbors C4-C7 (marked in skyblue). Since C0-C3 are in symmetric positions and C4-C7 are also in symmetric positions, there are 25 distinct combinations as shown in the figure.  
For this example, we set eCD=\SI{55}{\nano\meter} and pitch=\SI{90}{\nano\meter} (design spec. from the SK hynix high-density STT-MRAM design in \cite{Chung2016}).
It  can be seen that $H_{\mathrm{s\_inter}}^z$ reaches its lowest point (\SI{-16}{Oe}) when C0-C7 are all in 0 (P) state (i.e., NP\textsubscript{8}=0). In this case,  the magnetization in the FL of every aggressor cell is in parallel with that of the RL; together, they generate a stray field which is stronger enough to compensate the stray field from the HL. As the bit number of 1s increases, $H_{\mathrm{s\_inter}}^z$ increases;  it increases in a step of \SI{15}{Oe} with the number of 1s in direct neighbors and in a step of \SI{5}{Oe} with the number of 1s in the diagonal neighbors.  When  C0-C7  are all in 1 (AP) state (i.e.,  NP\textsubscript{8}=255),  $H_{\mathrm{s\_inter}}^z$  reaches the peak (\SI{64}{Oe}). Therefore, the maximum variation  in $H_{\mathrm{s\_inter}}^z$ among the 256 neighborhood patterns is \SI{80}{Oe} in this case. If the value is too large compared to the device coercivity ($H_{\mathrm c}$=\SI{2.2}{kOe} for the measured devices in this paper), it may result in a significant variation in the device performance.  To quantitatively evaluate the inter-cell magnetic coupling strength, we  defined  { \em inter-cell magnetic coupling factor} $\Psi$ as  the ratio of the maximum variation in $H_{\mathrm{s\_inter}}^z$ to $H_{\mathrm c}$. 
 $\Psi$ will be used as an indicator of inter-cell magnetic coupling strength  in the rest of this paper.
 
The $\Psi$ value varies with device size and array pitch, as shown in Fig. \ref{MC_modeling_results2}b.  In our simulations, we set the minimum pitch to 1.5$\times$eCD according to \cite{Nguyen2018} for high-density STT-MRAMs and the maximum pitch to \SI{200}{\nano\meter}, which is adopted by both Samsung and Intel \cite{Song2016,Wei2019a}. It can be seen that $\Psi\approx0$\%  at  pitch=\SI{200}{\nano\meter} for all three device sizes, indicating the inter-cell magnetic coupling is negligible due to the far distance between devices. As the pitch decreases, $\Psi$ increases gradually until reaching  a threshold point after which it goes up exponentially. For our devices,  $\Psi$=2\% (marked with the dashed line) can be considered as  the threshold point, where the array density is maximized with negligible inter-cell magnetic coupling. For a device with eCD=\SI{35}{\nano\meter}, this corresponds to pitch=$\sim$\SI{80}{\nano\meter}.

\section{Evaluation of  Magnetic Coupling Impacts}
\label{sec:MC_impact}
In this section, we evaluate the impact of magnetic coupling on the critical switching current  $I_{\mathrm{c}}$ and the  average switching time $t_{\mathrm{w}}$, using the proposed model.    Thereafter, we investigate the impact on the thermal stability factor $\Delta$ in a similar way.  Due to the space limitation, we will only present the simulation results for devices with eCD=\SI{35}{\nano\meter}.

%

\subsection{Impact on the Critical Switching Current}
\label{subsec:Ic}
Under the influence of stray field, $I_{\mathrm{c}}$ can be expressed as  follows \cite{Khvalkovskiy2013d}:
\vspace{-5pt}
\begin{equation}
I_{\mathrm{c}} (H_{\mathrm{stray}}^z) =\frac{1}{\eta } \frac{2\alpha e}{\hbar}  M_{\mathrm{s}} \cdot V \cdot H_{\mathrm{k}}\cdot(1\pm\frac{H_{\mathrm{stray}}^z}{H_{\mathrm k}}) ,  \label{eq:Ic} \\
\end{equation}
where $\eta$ is the STT efficiency, $\alpha$ the magnetic damping constant, $e$ the elementary charge, $\hbar$ the reduced Planck constant, $M_{\mathrm s}$ the saturation magnetization, $V$ the volume of the FL, $H_{\mathrm k}$  the magnetic anisotropy field. 
The sign   in the parentheses is  `$+$' for $I_{\mathrm{c}}$(P$\rightarrow$AP) and  `$-$'  for $I_{\mathrm{c}}$(AP$\rightarrow$P), given the definition of coordinates in this paper. 
In Equation~(\ref{eq:Ic}), $H_{\mathrm{stray}}^{z}=H_{\mathrm{s\_intra}}^{z}+H_{\mathrm{s\_inter}}^{z}$ can be calculated with our proposed magnetic coupling model taking into account both  intra-cell and  inter-cell stray fields, while $H_{\mathrm{k}}$  needs to be extracted from measurement data. the other parameters in the equation are measured at blanket stage before etch.
Since the switching points (i.e., $H_{\mathrm{sw\_p}}$ and $H_{\mathrm{sw\_n}}$ in Fig.~\ref{fig:Hs_characterization}a) are intrinsically stochastic,  we measured the R-H loop of the same device for 1000 cycles to obtain a statistical result of the switching probability at varying fields. With the technique proposed in \cite{Thomas2014}, we are able to extract $H_{\mathrm{k}}$ and $\Delta_0$ by performing curve fitting.  $\Delta_0$ is the intrinsic thermal stability factor without any stray field at the FL; it will be used in the next subsection.  By doing this for a large number of devices, we obtained  $\Delta_0=45.5$ and $H_{\mathrm{k}}=\SI{4646.8}{Oe}$  (both in median) for devices with eCD=\SI{35}{\nano\meter}.

Fig. \ref{MC_modeling_results2}c shows  the critical switching current $I_{\mathrm{c}}$ for C8 (for both P$\rightarrow$AP switching and AP$\rightarrow$P switching) at different pitches with respect to various stray fields. 
For isolated devices without any stray field (i.e., ideal case, $H_{\mathrm{stray}}^{z}=0$), the intrinsic $I_{\mathrm{c}}$  for the two switching directions is supposed to show no difference; $I_{\mathrm{c}}=\SI{57.2}{\micro\ampere}$. When taking into account the intra-cell stray field (i.e., $H_{\mathrm{stray}}^{z}=H_{\mathrm{s\_intra}}^{z}$), a static shift in $I_{\mathrm{c}}$ is introduced,  making 
$I_{\mathrm{c}}$(AP$\rightarrow$P)=\SI{61.7}{\micro\ampere} (i.e., 7\% above the intrinsic $I_{\mathrm{c}}$)  and $I_{\mathrm{c}}$(P$\rightarrow$AP)=\SI{52.8}{\micro\ampere} (i.e., 7\% below). When considering both intra-cell and inter-cell stray fields (i.e.,  $H_{\mathrm{stray}}^{z}=H_{\mathrm{s\_intra}}^{z}$+$H_{\mathrm{s\_inter}}^{z}$) for different neighborhood patterns NP\textsubscript{8}, the impact on $I_{\mathrm{c}}$ shows a clear dependence on the array pitch.  $I_{\mathrm{c}}$(AP$\rightarrow$P) becomes larger at smaller pitches when  NP\textsubscript{8}=0, while it shows an opposite trend when NP\textsubscript{8}=255. This indicates that the variation in $I_{\mathrm{c}}$(AP$\rightarrow$P) between different neighborhood  patterns increases as the pitch goes down. 
It can be seen that at pitch$\approx$\SI{80}{\nano\meter} (corresponding to $\Psi=2\%$), the variation is marginal. Similar observations can be seen on the P$\rightarrow$AP switching direction.



\subsection{Impact on the Average Switching Time}
\label{subsec:tw}
The average switching time  $t_{\mathrm w}$ in the presence of  $H_{\mathrm{stray}}^{z}$ in the precessional regime (namely, switched by the STT-effect) can be estimated using Sun's model  as follows \cite{Wu2019a}:
\begin{align}
 t_{\mathrm w}(H_{\mathrm{stray}}^{z}) &= (\frac {2} {C+\ln(\frac {\pi ^2\Delta }{4})} \cdot \frac{\mu_{\mathrm{B}}P}{e m(1+P^2)}\cdot I_{\mathrm{m}})^{-1}, \\
I_{\mathrm{m}}&=\frac{V_{\mathrm{p}}}{R(V_{\mathrm{p}})}- I_{\mathrm{c}}(H_{\mathrm{stray}}^{z})  .  \label{eq:tw_prec}
\end{align}
Here, $C$$\approx$$0.577$ is Euler's constant, $\mu_{\mathrm{B}}$ the Bohr magneton, $P$ the spin polarization,  $e$ the elementary
charge, and $m$ the FL magnetization. $V_{\mathrm{p}}$ is the voltage applied on the MTJ device to switch its state. $R({V_{\mathrm{p}}})$ is the resistance of the MTJ device as a function of  the applied voltage $V_{\mathrm{p}}$; it shows a non-linear dependence on $V_{\mathrm{p}}$ \cite{Wu2019a}.

\begin{figure*}[!t]
	\centering
	\includegraphics[width=1 \textwidth ]{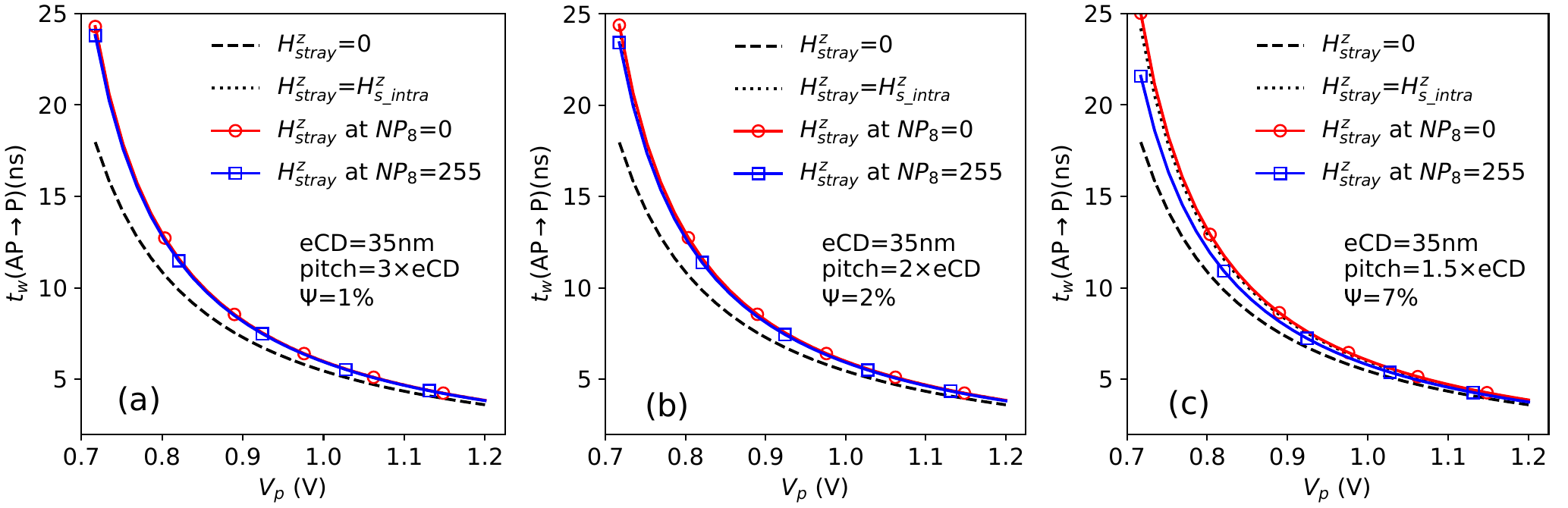}
	\caption{Impact of magnetic coupling on the voltage dependence of $t_{\mathrm{w}}$(AP$\rightarrow$P)  with eCD=\SI{35}{\nano\meter} at various pitches: (a) 3$\times$eCD, (b) 2$\times$eCD, and (c) 1.5$\times$eCD.}
	\label{fig:tw_vs_Vmtj}
\end{figure*}

Fig. \ref{fig:tw_vs_Vmtj}a-c shows the voltage dependence of the average switching time from AP state to P state ($t_{\mathrm{w}}$(AP$\rightarrow$P)) for MTJs with eCD=\SI{35}{\nano\meter} at pitch=3$\times$eCD, 2$\times$eCD, and 1.5$\times$eCD. Due to the space limitation, the simulation results of $t_{\mathrm{w}}$(P$\rightarrow$AP) are excluded. It can be seen that $t_{\mathrm{w}}$(AP$\rightarrow$P) becomes larger for MTJ devices in the presence of $H_{\mathrm{stray}}^z$ (solid lines), comparing to devices without any stray field (dashed lines). It is worth noting that the larger the voltage, the smaller the impact of the stray field on $t_{\mathrm{w}}$(AP$\rightarrow$P). However, an increase in the switching voltage $V_{\mathrm{p}}$ also results in more power consumption and a higher vulnerability to breakdown. In addition, when the pitch goes from 3$\times$eCD (Fig.~\ref{fig:tw_vs_Vmtj}a) to 2$\times$eCD (Fig.~\ref{fig:tw_vs_Vmtj}b), the inter-cell magnetic coupling factor $\Psi$ increases from 1\% to 2\% and the change in $t_{\mathrm{w}}$(AP$\rightarrow$P) is negligible. However, when the pitch goes down to 1.5$\times$eCD (Fig. \ref{fig:tw_vs_Vmtj}c),  $\Psi$ increases to 7\% and the variation in $t_{\mathrm{w}}$(AP$\rightarrow$P) between different NPs (i.e., $H_{\mathrm{s\_inter}}^{z}$) becomes very visible. For example, at a voltage of \SI{0.72}{\volt}, $t_{\mathrm{w}}$(AP$\rightarrow$P)  under NP\textsubscript{8}=0 is $\sim$\SI{4}{\nano\second} slower than  NP\textsubscript{8}=255, as shown in Fig. \ref{fig:tw_vs_Vmtj}c. This indicates that a  larger write margin (e.g., a longer pulse) is required to avoid write failure in the worst-case (i.e., NP\textsubscript{8}=0).


\begin{figure}[!t]
	\centering
	\includegraphics[width=0.5 \textwidth ]{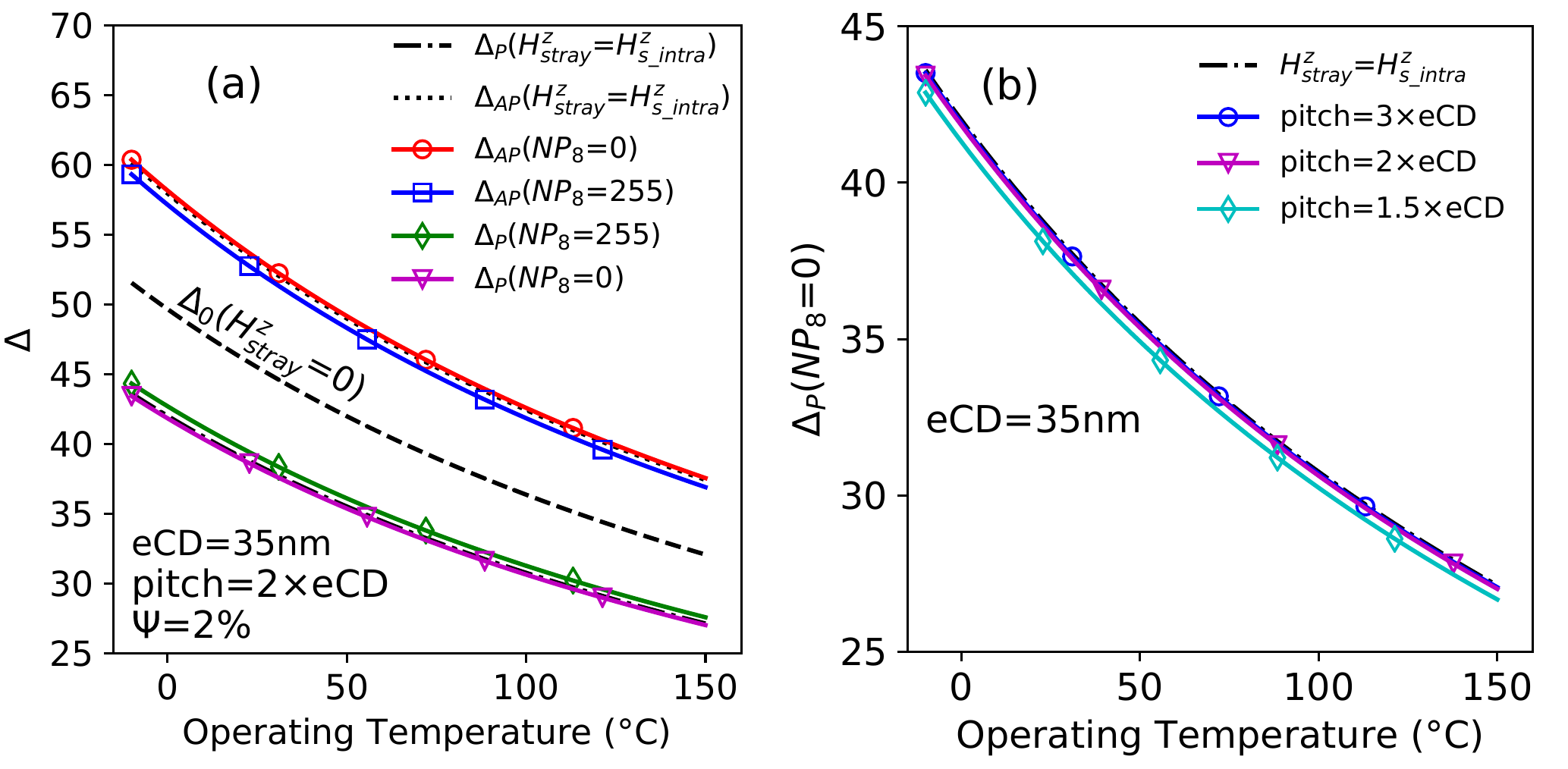}
	\caption{Impact of magnetic coupling on  $\Delta$  with  eCD=\SI{35}{\nano\meter} at: (a) pitch= 2$\times$eCD and (b) worst-case $\Delta$  for pitch=3$\times$eCD, 2$\times$eCD, and 1.5$\times$eCD. }
	\label{fig:Delta_vs_T}
\end{figure}

\subsection{Impact on the Thermal Stability Factor}
\label{subsec:Delta}
The intrinsic thermal stability factor $\Delta_0$ (without any stray field at the FL) of the MTJ device is given by \cite{Khvalkovskiy2013d}:
$\Delta_0 =\frac{H_{\mathrm{k}}M_{\mathrm{s}}V}{2k_BT}$,
where  $k_{\mathrm{B}}$ is the Boltzmann constant and $T$  is the absolute temperature. However, in the presence of stray fields, the thermal stability factor in AP state deviates from that in P state, i.e., $\Delta_{\mathrm{AP}}$$\neq$$\Delta_{\mathrm{P}}$. The $\Delta$ value in the presence of $H_{\mathrm{stray}}^z$ is given by \cite{Khvalkovskiy2013d}:
\begin{equation}
\Delta(H_{\mathrm{stray}}^z)   =\Delta_0(1\pm\frac{H_{\mathrm{stray}}^z}{H_{\mathrm{k}}})^2,   \label{eq:Delta_Hstray} 
\end{equation}
where the sign   in the parentheses is  `$+$' for $\Delta_{\mathrm{P}}$ and  `$-$'  for $\Delta_{\mathrm{AP}}$ for the devices considered in this paper.  $H_{\mathrm{stray}}^z$ can be calculated with our proposed magnetic coupling model, while $H_{\mathrm{k}}$ and $\Delta_0$ are extracted from measurement data.

Fig. \ref{fig:Delta_vs_T}a shows the  thermal stability factor $\Delta$ at varying temperature for eCD=\SI{35}{\nano\meter} and pitch=2$\times$eCD, corresponding to $\Psi=2\%$.  It can be seen that the intra-cell stray field $H_{\mathrm{s\_intra}}^{z}$ introduces a static shift in $\Delta_{\mathrm{AP}}$ and $\Delta_{\mathrm{P}}$; $\Delta_{\mathrm{AP}}$ is $\sim$30\% smaller than  $\Delta_{\mathrm{P}}$ comparing the dash-dotted line to the dotted one. The solid lines represent the thermal stability factors considering both intra-cell and inter-cell magnetic coupling. It can be seen that the MTJ device has the smallest $\Delta$ (highest vulnerability to a retention fault) when  the victim cell is in P state and all neighboring cells are also in P state (i.e., NP\textsubscript{8}=0). Fig. \ref{fig:Delta_vs_T}b compares the worst-case $\Delta$, i.e., $\Delta_{\mathrm{P}}$(NP\textsubscript{8}=0), at pitch=3$\times$eCD, 2$\times$eCD, and 1.5$\times$eCD.
One can observe that $\Delta_{\mathrm{P}}$(NP\textsubscript{8}=0) shows a marginal degradation when the array pitch goes down to 1.5$\times$eCD, in comparison to pitch=2$\times$eCD.
%
%
\section{ Conclusion}
\label{sec:conclusion}
Magnetic coupling including both intra- and inter-cell can be a critical design constraint for high-density STT-MRAM designs as the MTJ device scales down. Intra-cell magnetic coupling needs to be minimized by further device stack innovation,  as it leads to a significant bifurcation in the switching characteristics and retention  for AP$\rightarrow$P and P$\rightarrow$AP switching directions. The inter-cell magnetic coupling depends on the device size as well as the array pitch. When the pitch reaches at $\sim$2 times of device diameter (corresponding to $\Psi=2\%$), the array density is maximized with negligible impact on the device performance. However, more aggressive exploration in array density (e.g., pitch=1.5$\times$eCD) shows significant variations in average  switching time, especially at low switching voltages, caused by inter-cell magnetic coupling. Moreover, we also observed a marginal degradation of retention due to the increased inter-cell magnetic coupling. 
\section*{Acknowledgment}
We would like to acknowledge and thank the efforts of Sebastien Couet, Farrukh Yasin and Davide Crotti in fabricating the STT-MRAM devices and Nico Jossart for providing the SEM images. This work was performed as part of imec's industrial affiliation program on STT-MRAM devices.

\ifCLASSOPTIONcaptionsoff
  \newpage
\fi


\def\url#1{}
\bibliographystyle{IEEEtran}
\bibliography{Mybib,library}{}

\end{document}